# A note concerning Primary Source Knowledge

HM COLLINS, P GINSPARG, L REYES-GALINDO

**Introduction: Primary Source Knowledge and the arXiv automated filter**

We add a small increment to understanding the notion of 'Primary Source Knowledge' (PSK). PSK is knowledge that the non-expert and the citizen can acquire by assiduously reading the primary scientific journal literature without being embedded in the cultural life of the corresponding technical specialty (Collins and Evans 2007). This comes from exposing four papers to the automated computer filters used by the physics preprint server 'arXiv. These filters are used to flag papers in need of further review by human assessors before being promulgated on the server; papers not flagged by the algorithm are generally posted on arXiv without further review.[1] After the filtering, human moderators decide whether papers should be posted based on a relatively low bar of whether they are of "interest, relevance and value" to the research communities that populate arXiv.[2]

The purpose of the exercise was to test whether the filters are able to flag papers that, while technically adequate, are now considered 'outdated' and ignorable by a research community. As we discuss below, taking the filters as (generous) proxies for non-experts, the answer has important ramifications in terms of the usability of scientific information by non-experts, citizens, policymakers and other audiences who are outsiders to scientific research communities. The general argument is that if arXiv's sophisticated filters are unable to spot 'outdated' contributions that are ignored by the core members of a research community, one can envisage that outsiders to a research community would be unlikely to spot them either, allowing for the inadequate or uninformed use of otherwise 'legitimate' scientific knowledge. This would suggest constraints on the level of usability that scientific knowledge has for non-expert audiences, even in the face of the unrestricted availability of scientific repositories like arXiv.

**The arXiv filter efficiency**

The arXiv filters work in two steps, using well-known computer linguistic-analysis techniques in tandem that incorporate both content and stylistic analysis. Broadly speaking, the first implies looking at 'keywords' and 'content words' (and combinations of them) and matching the frequency of their occurrence to the patterns of an existing arXiv research community (Ginsparg et al 2003). If no match is found, the paper is flagged as suspect and in need of further (human) scrutiny, as arXiv's main categories are modelled to be outlets for existing research communities. The second filter ignores the technical content and looks purely at an author's writing style based on the most commonly used words, which are known to follow statistical laws and which Ginsparg has found to be an additional marker for membership to a scientific community (stylistic deviations, e.g. authors having English as a second language, are also taken into account). Similar instances of topic identification through latent semantic indexing (Deerwester et al 1990), latent Dirichlet allocation (Blei and Jordan 2003), topic modelling (Griffiths and Steyvers 2004) and probabilistic latent semantic analysis (Hofmann 1999) have been successfully carried out during the past three decades, but to the authors' best knowledge no other implementation of the magnitude of the arXiv filters exists and much less run in real time –using a corpus of over half a million scientific papers over the last decade. We believe that arXiv's filters represent the 'state of the art' as regards automated procedures for separating papers that fit with current scientific topics and styles of presentation from those that do not.

---

[1] Though human assessors may choose to examine papers that have not been automatically flagged.
[2] https://arxiv.org/help/moderation

To test the filter efficiency, a detailed analysis was carried out by Ginsparg that looked at the filtering success rate for the year July 2014-July 2015 by comparing how the filter pre-screening outcomes align to final human moderator decisions. Out of 100,904 total submissions to arXiv, the analysis concentrated on the 55,667 submissions to the physics category (55.2% of arXiv total). 537 submissions were rejected by arXiv for publication (0.98% of physics total) and the filters were successful in identifying 88% of these correctly as candidates for rejection – this figure excludes a number of submissions that the filters could not tackle for various technical reasons such as insufficient length, submission being in language other than English and other technical points. Ginsparg also produced statistics for papers accepted, but reclassified into the low-status 'general physics' category, which published 302 papers in this time period (0.54% of physics total). Again excluding submissions that could not be handled by the filters for technical reasons, 91% were appropriately flagged by the filters.

**arXiv as an Open Access source of Primary Source Knowledge**

Collins and Evans argue that Primary Source Knowledge can be misleading because to understand and make use of a scientific paper requires more than technical comprehension; it requires an understanding of whether the paper is taken seriously within the corresponding community of experts. Science is not just an aggregate of theories and experimental results; when there is dispute, it is often the informal, assessment of the credibility of results that lead to one conclusion rather than another. This social knowledge is not part of the paper itself and only rarely do hints of it appear in the published literature such as when there are series of 'Comments on X' and replies appear in technical journals. We should make clear that we are not here discussing the truth or otherwise of the claims made in the papers discussed or the competence of the procedures supporting them, we are discussing their *credibility* within the scientific community only; this, of course, is the only resource available to those who need to make immediate assessments of the value of disputed scientific claims, whereas science proper can wait as long as necessary for what comes to be counted as truth to be agreed upon.

There are, then, a subset of published papers that working scientists simply do not take seriously. Some of these will have been written by scientists who have never developed a reputation for good research and some by scientists who were once part of the mainstream – perhaps even research leaders – but whose ideas have lost credibility. For example, consider double Nobel-laureate Linus Pauling's later papers on the cancer-curing properties of high doses of vitamin-C. In spite of Pauling's huge reputation and track record, very few took his vitamin-C claims seriously. Here we try to demonstrate that it is not possible for citizens to separate papers that are taken seriously and those that are not taken seriously because citizens are not immersed in the cultural life of the corresponding specialty which is the only place that kind of understanding can be acquired. This understanding can be called 'Domain-Specific Discrimination' (Collins & Weinel 2011).

This has consequences. For example, in 1999 Mbeki spoke as follows to the second chamber of the South African parliament:

> There … exists a large volume of scientific literature alleging that, among other things, the toxicity of this drug [the anti-retroviral AZT] is such that it is in fact a danger to health. … To understand this matter better, I would urge the Honourable Members of the National Council to access the huge volume of literature on this matter available on the Internet, so that all of us can approach this issue from the same base of information.

Mbeki was advising parliamentarians to take advantage of Primary Source Knowledge so as to reach a decision on the safety of anti-retroviral drugs. The result, it has been said, was 10,000 avoidable deaths – avoidable because the papers which fed the internet controversy were no longer taken seriously within the expert community even though they were written by scientists who were once leaders in the relevant field. Experts immersed in the field would have known that the papers claiming that anti-retrovirals were unsafe and ineffective were no longer considered sound. They were 'past their sell-by date' (Weinel 2007).

Collins (2014) sets out the justifications scientists working with the Laser Interferometer Gravitational-Wave Observatory (LIGO) said could be deployed for rejecting, without further consideration, a specific paper in the field of physics. This paper claimed that LIGO's interferometers were incapable of detecting the phenomenon they were built to search for. There are many features of the paper that physicists could use to support the fact that they ignored it though it is almost certain that the paper would not have come to their attention at all had Collins not asked the question. It is likely that the assiduous citizen reading such a paper would not know how to assess it but there is a chance that they could learn to spot certain indicators such as high numbers of self-citations. But we now know that the paper in question would have been flagged by arXiv's automated algorithm which can recognise problems by looking for more subtle cues; perhaps one day citizens could learn to use such algorithms themselves so as to demarcate the credible from the non-credible, or such bibliometric indicators could be used to bolster filters to supersede arXiv's. The question remains, then, as to whether there are papers in the literature that are past their sell-by date or not credible for other reasons but are not detectable either by reference to the indicators mentioned by the scientists in Collins (2004) nor by automatic flagging by arXiv. Collins and Galindo each chose two potential papers of this type based on their experience. Collins's relevant experience is his long-term sociological study of the field of gravitational wave detection while Galindo has done physics graduate work on the 'Casimir Effect'. Ginsparg offered the papers selected by Collins and Galindo to arXiv's algorithm so that we could find out if such methods could ever empower the citizen.

**Four papers offered up to arXiv**

We began with: Weber, Joseph, and B. Radak. 1996. "Search for Correlations of Gamma-Ray Bursts with Gravitational-Radiation Antenna pulses." *Nuovo Cimento B* 111 (6): 687–92. This is a paper published by the founder of the field of gravitational wave detection with terrestrial detectors, Joseph Weber. Weber's detection claims had lost credibility around 1975 among the vast majority of the expert community and this paper was published long after that date. Towards the end of the 1990s Collins conducted an informal survey of the gravitational wave detection community and could find no-one who had even *read* the paper even though, if its claims had been correct, it would have been Nobel Prize-worthy. By 1996 the credibility of Weber's detection claims had fallen so far – the sell-by date was so far in the past – that the paper was simply ignored.[3] On the face of it, however, the paper was a perfectly respectable, even exciting, piece of work. Confirming this, Ginsparg found that arXiv's automated algorithm did *not* flag the paper even though it was nearly 20 years old at the time of the test. (We feared that stylistic changes in the field of physics would make it look odd but this turned out not to be the case.)

We therefore decided to try arXiv out on similar papers. Unfortunately this is not easy. We could not try out the papers by Pauling or those implicated in the Mbeki case because arXiv deals with physics not biology. Worse, the very claim that we are trying to support – that there is a class of rejected papers that can be recognised only by fully enculturated domain-experts – meant that there was no automated way of finding the right kind of paper. The right kind of paper had to be written competently by a competent specialist but refer to a matter, or make a claim, which had been rejected by the scientific community for reasons to do with credibility not glaring technical error. The Weber paper was picked out by Collins consequent on his long immersion and deep cultural understanding of the field of gravitational wave detection and he felt he could select another paper from the same field that he knew was not taken seriously by experts even thought it was widely publicised at the time.

This was a paper claiming to have detected gravitational waves emitted by an exploding star: Aglietta, M., G. Badion, G. Bologna, C. Castagnoli, A. Castellina, W. Fulgione, P. Galeotti, et al. 1989. "Analysis of the Data Recorded by the Mont Blanc Neutrino Detector and by the Maryland and Rome Gravitational-Wave Detectors during SN 1987A." *Il Nuovo Cimento* 12C (1): 75–103. This paper reported that two gravitational wave detectors, one in Rome and one in Maryland, had seen signals coincident with each other and roughly coincident with emissions of neutrinos from the 1987 supernova. Given the lack of sensitivity according to the accepted theory of the out-dated room-temperature resonant bars involved in the claimed detection (all the more advanced detectors were off air), as the authors themselves explain, the amount of energy emitted as GWs must have been equivalent to the total conversion of an 'abnormal' 2,400 solar masses. The authors, therefore, simply report their observations without interpretation though the implication is clear.[4] All terrestrial GW observations depend on analysis of coincident signals seen on two widely separated devices. In this case one belonged to an Italian group and one to Joe Weber – whose observational reports were, as explained, no longer trusted in the community. The credibility of the Italians was also badly eroded since they were known to be continuing supporters of Weber and had already promulgated results which were not believed. Furthermore, this paper had been rejected by *Physical Review*, where it was first submitted, though *Physical Review* then published a critique of the paper and then refused to publish a response to the critique.[5] The text of the SN1987a paper was not flagged by arXiv as having any "fringe' aspects though it would have received a second look so as to decide whether it belonged in the 'instrumentation and method' category or in 'high energy physics' so this is another case where one has to have a deep understanding of debates within the field to know how little credibility was given to the paper by the community.[6]

---

[3] See Collins (2004) for the full story.

[4] To speculate, it is possible that they were encouraged by the fact that they knew there were maverick theories of detector sensitivity floating around – championed by Joe Weber – that, if accepted, would make the result more believable (these alternative theories are not mentioned in the paper).

[5] The events need to be understood in the context of battles over the funding of the Laser Interferometer Gravitational-Wave Observatory (LIGO) which were at their height – See Collins, 2004, pps 372-380 for the remarkable story – but there were many aspects of the paper which would have prevented it being believed even outside of the funding battle.

The next two papers put to the test were selected by Galindo. The first was Mostepanenko, V. M. 2015. "How to confirm and exclude different models of material properties in the Casimir effect". *Journal of Physics: Condensed Matter*, 27(21), 214013.[7] The author is a highly respected theoretician in the field of Casimir forces. This is a paper referring to a heated controversy that appeared in the early 2000s regarding the ability to carry out precision measurements of Casimir forces. The controversy was around the use of two well-known models of metal surfaces that were being used to compare phenomenological theory with real experimental data in order to make precision claims. At that time, Reyes-Galindo was a young researcher in the field within a group that specialised on these topics. Had the measurements been able to reach the precision publicly claimed by Mostepanenko and collaborators, these Casimir experiments could have had enormous impact in looking for extra-dimension physics and in applications to nanotechnology. However, other experimental and theoretical groups strongly resisted the theoretical interpretation and the precision claims. After a decade of acrimonious back and forth argument with the leading experts talking past each other, a new directions of research started to appear in the field in the mid 2000s that looked for alternative interpretations. By 2011, when Reyes-Galindo attended his last major Casimir meeting while still working as a physicist, no conclusive argument had been found for accepting or rejecting either model.[8] Though unresolved, the controversy by that time had fizzled out of importance as younger researchers looked in other directions. In the 2015 paper Mostepanenko rehashes the arguments from the controversy's early days. Asked to comment, an expert in Casimir forces considered that these type of papers "meet those characteristics, of people in the know looking at them and saying, 'oh no, not this *again*!' Mostepanenko's latter papers don't shed any new light on these subjects. But they are technically correct."[9]

The fourth paper is from the areas of physics that Reyes-Galindo worked in when completing his undergraduate degree. Reyes-Galindo co-authored a paper on acoustic analogies to the Casimir effect in which a critical quantity, 'acoustic radiation pressure', had to be precisely defined to carry out the theoretical analysis. In the first submitted version of the paper the authors had used the definition of the acoustic pressure put forth in a paper by Beyer (1978) based on an analysis of another quantity called the 'stress tensor'. The reference was approved by the senior co-author who had published several acoustics papers because: a) it was published in a highly ranked acoustics journal; b) had a significant number of citations; c) included an interesting historical discussion of the radiation pressure concept. The co-authored paper was accepted, but with a caveat: in addition to minor corrections recommended by two reviewers, one of them commented in the referee report:

---

[6] Sociological data is rarely completely 'clean' and it has to be said that the journal in which both of these gravitational wave papers were published, *Il Nuovo Cimento*, is not considered to be a high prestige journal among the gravitational wave physics community. Nevertheless, its reputation is not such as to enable one to advise citizens that nothing published in this journal is to be taken seriously.
[7] Also arXiv:1411.4548 [quant-ph], 17 November 2014.
[8] This was the last of the bi-annual QFEXT (Quantum Field Theory under External Conditions) conference series, which were the most important Casimir force meetings. Reyes-Galindo followed the development of the controversy as a participant observer at QFEXT'05 (Barcelona), QFEXT'09 (Oklahoma) and QFEXT'11 (Benasque). See L. Reyes-Galindo, Controversias en el efecto Casimir, Unpublished MPhil thesis, National Autonomous University of Mexico, 2007.
[9] It has to be said that the reputation of this work was still high among a group of physicists.

> "[Beyer (1978)] may not be the best reference for the acoustic radiation pressure, Eq (11). This article dealt with a silly polemic that surfaced in the 70's over an idiotic misunderstanding of radiation pressure in one-dimension (see Lord Rayleigh, Philos. Mag. 3, 338-346 (1902); Lord Rayleigh, Philos. Mag. 10, 364-374 (1905)). Using reference [this reference] may just be an invitation to renew this silliness."

Although Reyes-Galindo and colleagues had used a reference that appeared legitimate even to a physicist with publications in the field, a core member of the acoustics community had flagged it as a suspect paper. Once the minor changes suggested by the reference had been made and the offending reference substituted for the recommended literature, the paper was accepted for publications in three days. Beyer (1978) was run through the arXiv filter and, were it submitted today, would be automatically accepted as a 'normal' paper.[10]

**Citation patterns to papers without credibility**

The authors wondered if citation patterns to papers could act as a proxy for quality. The first problem for such a use of citations is that these can have a long latency – the citation history of a paper is still forming many years after publication. Thus, we cannot analyse the citation pattern of the Mostapanenko paper because it was published only in 2015 and so has only five preprint citations: two self-citations, one citation from close collaborators, and one citation highly critical of the results. Even for older papers lack of citations is not a reliable indicator of eccentricity because a high proportion of published papers are never cited and only a small subset of these uncited works will be lacking in *credibility* as opposed to simply lacking in interest or consequence.

In any case citation patterns to papers without credibility vary hugely.[11] For this work we carried out a rapid analysis of non-citation rates in high energy physics (HEP) using the inSPIRE database, which is historically a major curated bibliographic site for HEP and unlike major commercial databases includes up-to-date, full bibliographic arXiv data. Discarding self-citations, we found that out of 58,812 arXiv e-prints in the *hep-th* (theory) category that have been published in journals, approximately 11% remain uncited; for 9,462 in *hep-ex* (experiment) 6%; for 6,347 in *hep-lat* (lattice) 8%; for 61,705 in *hep-ph* (phenomenology) 8%; this gives a snapshot of HEP non-citation patterns as radically different from 'science and engineering' ones in general. Consider the notorious, 1998, paper in *The Lancet* by Andrew Wakefield and colleagues in which evidence was presented that could be interpreted as supporting the idea of a link between autism and mumps, measles and rubella (MMR) vaccine. This paper, currently marked as 'RETRACTED' on Google Scholar, had received 2142 citations as at 6 August 2015 more than 30 of them in 2015. Web of Science records 953 citations, 25 in 2015. This paper was unusual, however, in that the fact that it was the subject of controversy became widely known outside specialist circles. Much more unusual for a scientific paper was that the worthlessness of the claim linking autism to MMR, which was what gave rise to the public discussion, was obvious after even the most casual scrutiny: no evidence for a link to MMR was made in the paper – this being a subsequent interpretation – and the reported result from the tiny sample bore only on a possible link between measles virus in the gut and autism; single-shot measles vaccination was, nevertheless, recommended. This minimal level of scrutiny proved to be beyond the capacity of most journalists and the public at large. The other papers we discuss here are all reasonable in terms of scientific coherence and one would not expect a non-expert reader to be able to find flaws in their evidence or the way conclusions were drawn. Indeed, the point is that even expert readers would not find flaws beyond the normal cut and thrust of academic debate unless they knew the history of their field. But if we look at the papers we have discussed above, we still find considerable variation in citation patterns.

---

[10] The low reputation of the Beyer paper assigned here turns on the opinion of one referee but even if this example is treated as unreliable it shows how these things can work in principle.

[11] In fact, citation patterns vary tremendously across disciplines and in time. Using data from Web of Science, Larivière et al (2009) found that approximately 20% of papers in medicine, 40% in science and engineering, 45% in the social sciences and a whopping 95% in the humanities remain uncited 5 years after publication,. These statistics hold for papers published within the last half century, but Larivière et al also show that non-citation patterns vary greatly across disciplinary fields and also depend on the timeframes looked at. Other more limited studies also highlight the variability of citation patterns. Weale et al (2004) have argued for a correlation between journal impact factor and non-citation rates in the fields of immunology and surgery. Hu & Wu (2014) have focused on six individual journals across various fields and likewise find high variability in non-citation patters.

Thus, the SN1987a paper has been cited 67 times (as at 06/08/15) according to Google Scholar, though less than 10 times in the last five years and, for reasons we do not understand, these later citations are nearly always by Russians. We know, of course, that at least one of the 67 citations was from an article whose authors wrote only so as to reduce the wider credibility of the findings which were perceived as potentially damaging to LIGOs funding prospects.[12] Web of Science records 42 citations.

The paradigm case, Weber's 1996 paper, has been cited eight times as at 6 August 2015 according to Google Scholar and 7 according to Web of Science. Three of these are Collins's published discussion of the paper, one is by a journalist who wrote a popular book on GW, one is from Virginia Trimble, Weber's loyal wife, and three are the more 'normal' kind of citation.

Beyer's stress tensor paper has received 93 citations to-date according to Google scholar and 74 according to Web of Science. The citation rate has also been consistent from its publication to 2014, indicating that while the paper may actually be a useful reference today, the polemic raised by the paper in its original context is inaccessible to an outsider.

Thus citation rates by themselves, though they may give hints about a paper's reputation, cannot indicate credibility within an expert community. Absence of citations may give some support to other kinds of argument, however. A second paper, re-analysing and reinforcing the SN1987a paper – P. Galeotti, G.V. Pallottino, G. Pizzella, 'SN1987A: Revisiting the Data and the Correlation between Neutrino and Gravitational Detectors', was promulgated on arXiv in 2008 but has been cited only once according to inSPIRE, in a footnote in a PhD thesis.[13] This certainly suggests that these, on-the-face-of-it, stunningly important claims, were no more credible ten years after they were first promulgated and also that this lack of credibility was still invisible to arXiv's automated filters.[14]

It has to be stressed again that nothing reported here is meant to shed any doubt or otherwise on the scientific integrity or competence of the authors of the papers examined, only the credibility of the papers within the scientific community to which they were addressed. We are sure that the authors of the papers are confident that the claims therein have the potential to

---

[12] See Collins (2004).

[13] arXiv preprint arXiv:0810.3759 . Web of Science does not record arXiv citations.

[14] Though the paper was only published in a conference proceedings, which are rarely cited by physicists, the result would have been significant enough not have been ignored by the GW community if taken at face value.

be vindicated at some time in the future should the science in question come to be better understood.[15]

Conclusion

When rapid decisions have to be made about the scientific truth of the matter there is no better resource than the current consensus of the scientific community; this is not necessarily the truth of the matter as truth can change as scientific understanding grows but the current consensus is the *best* available technical input to technological debates in the public domain. What we have tried to show here is that embedding in the community of domain-experts may be the only way to learn Domain-Specific Discrimination and thus come to understand the current technical consensus and that reading the literature, even the primary literature, is not a substitute. Furthermore, our analysis shows that even the most currently accurate automatized tools such as arXiv's sophisticated filters cannot replace experts' judgements regarding credibility and consensus. This naturally does not in principle rule out the possibility of future advances to carry out such judgements and systems with access to more features will likely be more discriminating; work on such a system might even help to characterize and more accurately pinpoint some constituents of Primary Source Knowledge.

---

[15] For a discussion of such 'Sleeping Beauty' papers, see Ke et al (2015).